\documentclass[reprint, showpacs, amsmath, amssymb, aip, rsi]{revtex4-1}

\usepackage[latin1]{inputenc}
\usepackage[english]{babel}
\usepackage{graphicx}
\usepackage[]{color}

\usepackage[unicode=true, bookmarks=true, bookmarksnumbered=false, bookmarksopen=false, breaklinks=false, pdfborder={0 0 1}, backref=false, colorlinks=false]{hyperref}
\hypersetup{pdftitle={no title}, pdfauthor={S. Probst, F. B. Song, M. Weides, P. A. Bushev, A. V. Ustinov}, pdfkeywords={superconducting resonators, fitting methods}}
\usepackage[sort&compress]{natbib}

\begin{document}
\title{Efficient and robust analysis of complex scattering data under noise in microwave resonators}

\author{S.~Probst}
\email{sebastian.probst@kit.edu}
\affiliation{Physikalisches Institut, Karlsruhe Institute of Technology, D-76128 Karlsruhe, Germany}
\author{F.~B.~Song}
\affiliation{Physikalisches Institut, Karlsruhe Institute of Technology, D-76128 Karlsruhe, Germany}
\affiliation{The 10th Institute of Chinese Electronic Technology Corporation, Chengdu 610036, China}
\author{P.~A.~Bushev}
\affiliation{Experimentalphysik, Universit\"{a}t des Saarlandes, D-66123 Saarbr\"{u}cken, Germany}
\author{A.~V.~Ustinov}
\affiliation{Physikalisches Institut, Karlsruhe Institute of Technology, D-76128 Karlsruhe, Germany}
\affiliation{Laboratory of Superconducting Metamaterials, National University of Science and Technology ``MISIS'', Moscow 119049, Russia}
\author{M.~Weides}
\affiliation{Physikalisches Institut, Karlsruhe Institute of Technology, D-76128 Karlsruhe, Germany}
\affiliation{Institut f\"{u}r Physik, Johannes Gutenberg-Universit\"{a}t Mainz, D-55099 Mainz, Germany}

\date{\today}

\begin{abstract}
Superconducting microwave resonators are reliable circuits widely used for detection and as test devices for material research. A reliable determination of their external and internal quality factors is crucial for many modern applications, which either require fast measurements or operate in the single photon regime with small signal to noise ratios. Here, we use the circle fit technique with diameter correction and provide a step by step guide for implementing an algorithm for robust fitting and calibration of complex resonator scattering data in the presence of noise. The speedup and robustness of the analysis are achieved by employing an algebraic rather than an iterative fit technique for the resonance circle.
\end{abstract}


\maketitle

\section{Introduction}
The evaluation of the scattering parameters of resonators is crucial for a wide range of research directions. Resonators made out of superconductors provide low internal loss and are used to detect and study physical systems down to the single photon regime. For instance, in circuit quantum electrodynamics \cite{Blais_04} resonators are used for dispersive readout or coupling to quantum objects such as qubits \cite{Jerger_2012, Weides_Transmon, Braumueller2014} or spin systems \cite{Probst_PRL13, Kubo2010, Kubo2011, Shuster2010}. Thus, it is crucial to understand the physics behind these resonators and, moreover, to be able to precisely determine their characteristic quantities. Fitting methods for determining parameters of a resonator such as its resonance frequency $f_r$, coupling strength $Q_c$ or internal quality factor $Q_i$ were studied in several publications before \cite{Petersan_JAP98,GaoPhD2008,Khalil_12,Deng2013,Megrant_APL12,wallraff2008}. However, the high-noise environment as frequently found in single or few photon scattering experiments imposes a considerable challenge. In this paper, we first summarize the resonator properties and the conventional fitting procedures. Next, we provide a step-by-step explanation of the circle fit method, and discuss how to subtract the microwave environment of the resonator (cable length, signal amplification, etc.). The robustness of this fitting method is illustrated by adding noise to the signal.  \\

A resonator is characterized by its resonance frequency $f_r$ and quality factor $Q$. The quality factor is defined by the ratio of the energy stored in the resonator to the average energy loss per cycle times $2\pi$. There are usually different energy relaxation paths, which contribute to the quality factor. In general, internal and coupling losses are distinguished, which are described by the internal $Q_i$ and coupling or external $Q_c$ quality factors, respectively. The loaded or total quality factor can be obtained by adding up the reciprocal values of these two contributing quality factors \cite{Pozar_book,Khalil_12} $Q_l^{-1} = Q_i^{-1} + \text{Re}\left\{ Q_c^{-1} \right\}$. As one can see, the coupling quality factor can be a complex number, which will be addressed later.
\begin{figure}[ht!]%
\centering
\includegraphics[width=8.6cm]{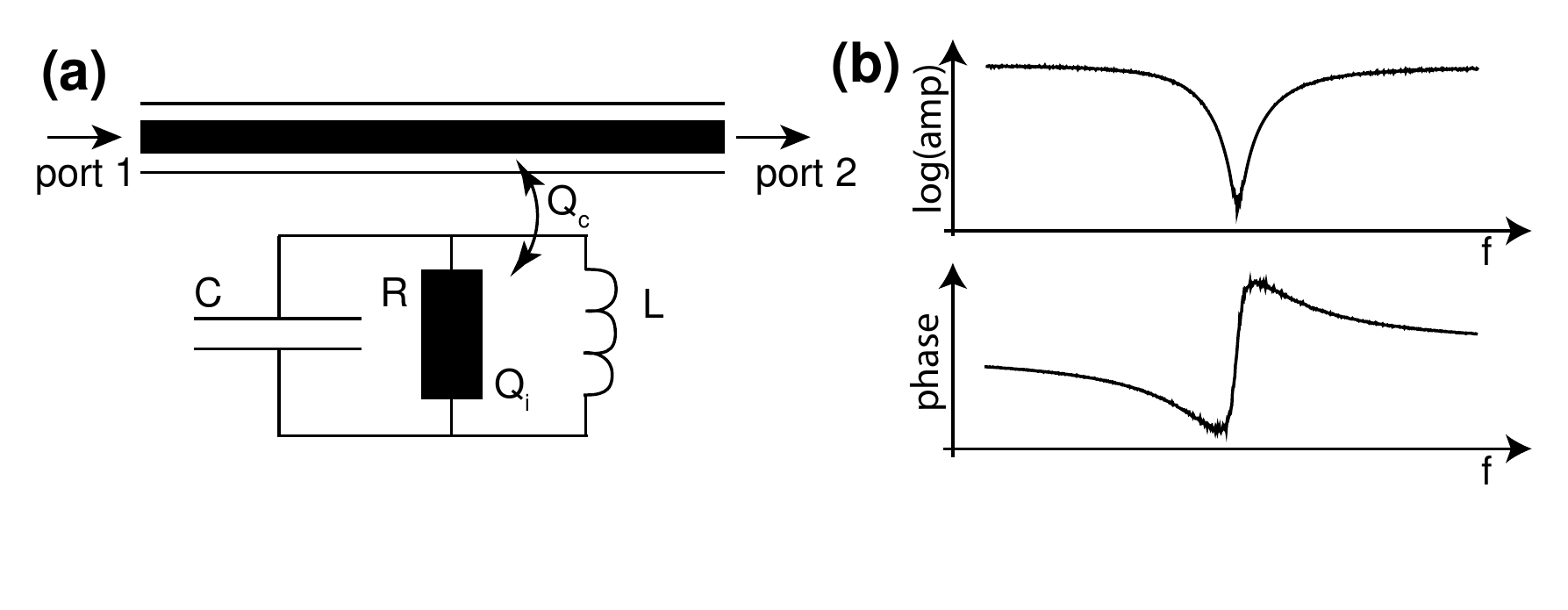}
\caption{(color online) (a) An LCR resonator coupled to transmission line. (b) Simulated transmitted amplitude and phase signal as a function of frequency with noise.}%
\label{fig:setup}%
\end{figure}
Resonators are usually measured either in reflection or transmission. In contrast to a reflection measurement, it is impossible to directly determine the internal losses of the resonator by probing the resonator's transmission due to the missing reference baseline. Therefore, the reflection measurement is much more appealing, and can be improved to choosing a so called notch type geometry, where the resonator is coupled to a transmission line. Figure \ref{fig:setup}(a) shows such a geometry, which allows for frequency division multiplexed readout of multiple resonators \cite{Jerger_2012,Wuensch2011,martinis2012,Vissers2012}. The chip is measured in transmission $S_{21}$ and the resonance appears as a dip in the spectrum. Figure \ref{fig:setup}(b) illustrates the complex transmission through such a notch type resonator chip by plotting the amplitude and phase of the transmitted signal as a function of the probe frequency.

The shape of the resonance dip can be asymmetric and there is an ongoing discussion on whether impedance mismatches in proximity to the resonator are responsible \cite{Khalil_12} or reflections between the input and output ports \cite{Deng2013}. Here, we follow the model by Khalil \textit{et al.}~\cite{Khalil_12} and use a complex $Q_c=|Q_c|\exp(-i\phi)$, which takes impedance mismatches into account quantified by $\phi$. $Q_c$ is a measure for the coupling to the external stimulus and readout circuit, but in most cases the internal quality factor $Q_i$ is of interest because it quantifies internal losses. Usually, these are caused by coupling to two level fluctuators, which occur due to impurities in the superconductor or substrate \cite{Gao_APL07,GaoAPL08,Gao_APL08b,Wang2009}. Another important application are hybrid systems, e.g. for quantum memory research. Here, the change of $Q_i$ is a direct measure for the coupling between resonator and spin ensemble \cite{Bushev2011,Shuster2010,Probst2014b}.\\

\begin{figure}[ht!]%
\centering
\includegraphics[width=8.6cm]{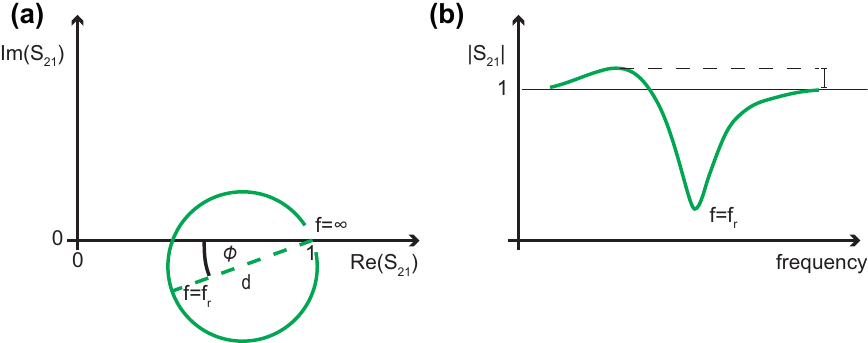}
\caption{(color online) (a) The resonance circle in its canonical position. The circle always intersects the real axis at 1 for $f\rightarrow \pm\infty$ independent of the other parameters, and the resonance $f=f_r$ is always located at the opposite side. A tilt by $\phi$ indicates an impedance mismatch and the diameter $d$ is equal to $Q_l/|Q_c|$. (b) Sketch of the corresponding amplitude signal. The rotation of the circle by $\phi$ causes an asymmetry in the amplitude signal. Note, that the signal above 1 is just an artifact of the standard normalization, which sets the off-resonant point to 1, see (a).}%
\label{fig:idealresonator}%
\end{figure}

There are a lot of different methods for extracting the resonator parameters from measured data. Petersan \textit{et al.} were the first to provide a quantitative comparison of the different methods \cite{Petersan_JAP98} at the time. A more recent method is presented by Deng \textit{et al.}~\cite{Deng2013}. Since vector network analyzers (VNAs) can measure the full complex transmission, modern methods make use of the entire complex scattering data instead of solely the power or amplitude. A general model derives the complex $S_{21}$ scattering coefficient of a notch type resonator as \cite{Khalil_12,GaoPhD2008}:
\begin{equation}
S_{21}^\text{notch}(f) = \underbrace{\vphantom{\frac{Q_l e^{i\phi}}{1+2iQ_l\left(f/f_r -1  \right)}}  a e^{i\alpha} e^{-2\pi i f \tau}}_{\text{environment}} \underbrace{\left[ 1 - \frac{\left(Q_l/ |Q_c|\right) ~ e^{i\phi}}{1+2iQ_l\left(f/f_r -1  \right)}  \right]}_{\text{ideal resonator}}\,.
\label{eq:S21model}
\end{equation}
The equation is split into two parts describing the ideal resonator and the environment. Here, $f$ denotes the probe frequency, $f_r$ the resonance frequency, $Q_l$ the loaded and $|Q_c|$ the absolute value of the coupling quality factor and $\phi$ quantifies the impedance mismatch. The environment is accounted for by adding an additional amplitude $a$, a phase shift $\alpha$ and the electronic delay $\tau$ caused by the length of the cable and finite speed of light. Figure \ref{fig:idealresonator} shows a sketch of the ideal resonator transmission, where its imaginary part $\Im\{S_{21}\}$ is plotted versus its real part $\Re\{S_{21}\}$. The diameter of this so-called resonance circle corresponds to $d=Q_l/|Q_c|$. The circle intersects the real axis at unity, which corresponds to a probe frequency $f\rightarrow \pm\infty$, the so-called off-resonant point $P$. On the other side of this point, one finds the resonator at its resonance $f=f_r$. Impedance mismatches may cause a rotation of the circle by an angle $\phi$, where the center of the rotation is on the real axis at unity, see Ref.~[\onlinecite{Khalil_12}] for further details.

The advantage of the notch type configuration becomes obvious by looking at a transmission measurement of a resonator. If one recalls the basic model of the lumped element LC circuit and its transmission function, one finds accordingly\cite{Pozar_book,Petersan_JAP98}:
\begin{equation}
S_{21}^\text{trans}(f) = a e^{i\alpha} e^{-2\pi i f \tau} \left[ \frac{\left(Q_l/ |Q_c|\right) ~ e^{i\phi}}{1+2iQ_l\left(f/f_r -1  \right)}  \right]\,.
\label{eq:S21model_transmission}
\end{equation}
However, the problem here is that with pure $S_{21}$ data, one cannot deduce the internal and coupling quality factors as well as the impedance mismatch. A quick way to see that is the following: The diameter of the ideal resonator's resonance circle is $Q_l/ |Q_c|$, but in an uncalibrated measurement, it is masked by an additional arbitrary amplitude $a$ resulting from the attenuation and gain in the measurement system, such that one measures $a\cdot Q_l/ |Q_c|$. It is possible to extract the value of $Q_l$, e.g. by fitting the phase, but this procedure is not sufficient to solve for $Q_c$ and, therefore, for $Q_i$. In other words, compared to the notch measurement, a reference baseline is missing. Since the off-resonant point $f\rightarrow \pm \infty$ is centered at the origin, one cannot distinguish between $\alpha$ and $\phi_0$, too. Different methods for fitting transmission resonators are discussed in Ref.~[\onlinecite{Petersan_JAP98}]. So, we will restrict ourselves to the discussion of notch type resonators, which have nowadays gained great importance, e.g., for multiplexed readout of qubits and microwave kinetic inductance detectors (MKIDs).

\section{Algebraic fit of the resonance circle}
The core task is to efficiently fit a circle in order to determine the diameter $d=Q_l/|Q_c|$. An algebraic method is preferable, because it allows to obtain the solution in a single run instead of iterative techniques, which suffer from the problem of choosing suitable initial parameters for reliable convergence. Here, we follow a paper by Chernov and Lesort, which compares several fitting methods for circles \cite{chernovlesort2005}. The advantage of this algebraic fit is that it needs neither any start parameter nor any iterations, and thus provides a fast and reliable tool. Especially, in the presence of heavy noise, this method is by far superior because it always gives a result and the time to compute that result is independent of the signal-to-noise ratio. Usually, algebraic solutions outperform iterative approaches. A useful parametrization of a circle is the following\cite{chernovlesort2005}
\begin{equation}
A(x^2+y^2)+Bx+Cy+D = 0\,.
\label{eq:circlepara}
\end{equation}
This over-parametrizes the problem and an additional constraint is needed:
\begin{equation}
B^2+C^2-4AD = 1\,.
\label{eq:constraint}
\end{equation}
This parametrization is powerful because it is more general and has no singularities. For $A=0$ one gets a line and $A>0$ always gives a circle. For convenience, we define $z=(x^2+y^2)$ and assume that our $n$ data points $(x_i,y_i)$ approximately form a circle: $Az_i+Bx_i+Cy_i+D \approx 0$. The function to be minimized reads as follows \cite{chernovlesort2005}:
\begin{equation}
\mathcal{F}(A,B,C,D) = \sum_{i=1}^n{\left( Az_i+Bx_i+Cy_i+D \right)^2}\,.
\label{eq:minfunc}
\end{equation}
$\mathcal{F}$ can be written in matrix form $\mathcal{F}=\textbf{A}^T M \textbf{A}$ with $\textbf{A} = (A,B,C,D)^T$. M is the matrix of the moments, e.g., $M_{xx} = \sum_{i=1}^n{x_i^2}$, $M_{xy}=\sum_{i=1}^n{x_i y_i}$.
\begin{equation}
M = \left(
\begin{array}{cccc}
	M_{zz} & M_{xz} & M_{yz} & M_{z}\\
	M_{xz} & M_{xx} & M_{xy} & M_{x}\\
	M_{yz} & M_{xy} & M_{yy} & M_{y}\\
	M_{z} & M_{x} & M_{y} & n\\
\end{array} \right)\,.
\label{eq:matrixofmoments}
\end{equation}
By analogy, the constraint from Eq.~(\ref{eq:constraint}) reads $\textbf{A}^T B \textbf{A}=1$ with matrix B defined as
\begin{equation}
B= \left(
\begin{array}{cccc}
	0 & 0 & 0 & -2\\
	0 & 1 & 0 & 0\\
	0 & 0 & 1 & 0\\
	-2 & 0 & 0 & 0\\
\end{array} \right)\,.
\label{eq:Bmatrix}
\end{equation}
The minimization is done using a Lagrange multiplier $\eta$
\begin{equation}
\mathcal{F^*} = \textbf{A}^T M \textbf{A} - \eta \left(\textbf{A}^T B \textbf{A} - 1 \right)\,.
\label{eq:lagrange}
\end{equation}
Differentiating with respect to \textbf{A} gives $M \textbf{A} = \eta B \textbf{A}$. The eigenvalues can be obtained by solving the characteristic polynomial $\xi(\eta)=\det\left(M-\eta B\right) = 0$. There is only one eigenvalue $\eta^*$ which minimizes $\mathcal{F^*}$. It has been shown that $\eta^*$ is the smallest non-negative eigenvalue. Furthermore, the characteristic polynomial is decreasing and concave up between 0 and $\eta^*$, see Ref.~[\onlinecite{chernovlesort2005}]. Thus, Newton's method with starting value 0 will certainly converge towards $\eta^*$.\\

For a practical implementation, the following steps are necessary:
\begin{enumerate}
	\item calculate all moments $M_{ij}$,
	\item find $\eta^*$ by solving the characteristic polynomial $\xi(\eta)$ using Newton's method with starting value $\eta=0$,
	\item find the corresponding eigenvector $\textbf{A}^*$ of $\eta^*$ using standard numerical algorithms.
\end{enumerate}

The radius and center coordinates of the circle can be extracted from $\textbf{A}^*$ by the following relations
\begin{eqnarray}
x_c &=& - \frac{B}{2A}\,, \\
y_c &=& - \frac{C}{2A}\,, \\
r_0 &=& \frac{1}{2|A|} \sqrt{B^2+C^2-4AD} = \frac{1}{2|A|}\,.
\label{eq:rxyrelations}
\end{eqnarray}
The relation for the radius $r_0$ is simplified because the expression in the discriminant is identical to our constraint in Eq.~(\ref{eq:constraint}), which was set to unity.\\

\section{Influence of the environment}
\begin{figure}[ht!]%
\centering
\includegraphics[width=8.6cm]{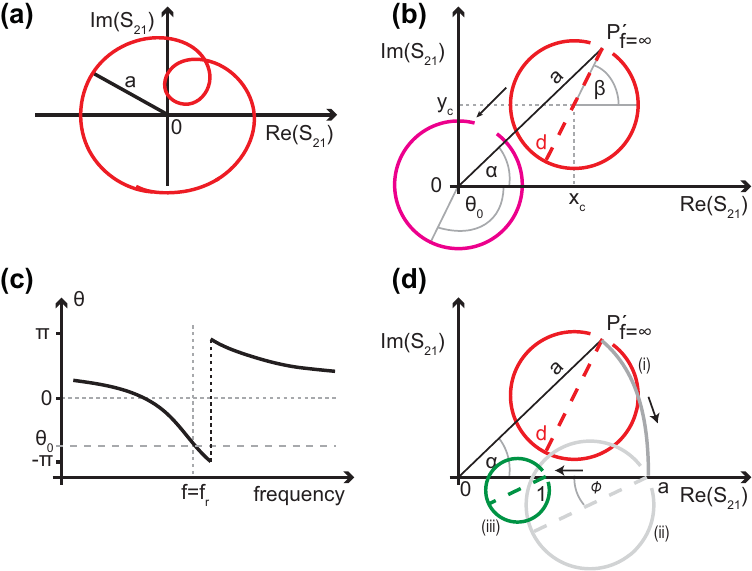}
\caption{(color online) (a) Uncalibrated transmission data of a notch resonator. The cable delay  causes a circular distortion of the ideal resonance circle. The radius $a$ of this distortion represents the system's attenuation/gain. The cable delay is removed by the procedure described in the text. (b) An algebraic fit determines the circle's diameter $d$ and its center position $(x_c,y_c)$. In order to determine the remaining prefactors $a$ and $\alpha$, the circle has to be translated to the origin. (c) A phase fit of the translated circle determines $\theta_0$, which yields $\beta$. This gives, together with the center and diameter of the circle, the coordinates of the off-resonant point $P'$ (see also (b)). (d) The final operation transforms the circle into its canonical position such that $P'\rightarrow 1$.}%
\label{fig:normalization}%
\end{figure}

$S_{21}$ data from an experiment differ from the idealized resonator model. The details of the experimental environment influence the measured $S_{21}$ data; Cable damping changes the amplitude $a$, the finite length of the cable causes additional rotations of the phase while scanning the frequency (cable delay $\tau$) and finally, the initial phase $\alpha$ might differ from 0. These influences can be accounted for by carefully calibrating the VNA's output signal. However, it is nontrivial to do this accurately, especially in the presence of an impedance mismatch. Here, an alternative way is chosen and these quantities are fitted as well.
Figure \ref{fig:normalization}(a) shows the uncalibrated data in the complex plane. The elimination of the prefactors in Eq.~(\ref{eq:S21model}) is done in two steps. First, a rough estimate of the cable delay is needed. The cable delay tilts the phase signal by a slope $2\pi \tau$ and in order to get a rough estimate, it is sufficient to fit a linear function to the phase signal. Without any cable delay, the resonance looks like a circle in the complex plane but the cable delay deforms this circle to a loop like curve. In the following, we perform a non-linear least square fit on this curve. The fit parameter is the cable delay and the error function to be minimized is the deviation from the ideal circular shape $\chi^2 = \sum_{i=1}^n{\left\{r_0^2-[(x_i-x_c)^2+(y_i-y_c)^2]\right\}}$. Here, $r_0$, $x_c$ and $y_c$ are the radius and the center coordinates obtained by performing the algebraic circle fit to the data. We emphasize that this procedure is only possible due to the robustness of the algebraic fit.\\
After having determined the cable delay $\tau$, the modified $S_{21}$ data already looks like a circle, but it still differs from the ideal case by an affine transformation, namely $ae^{i\alpha}$, see Fig.~\ref{fig:normalization}(b). Note, that $a$ accounts for additional attenuation/amplification present in the setup, however, it does neglect frequency dependent damping, e.g., of microwave cables. In most cases, the resonance is narrow and this effect is negligible.\\
In order to determine $a$ and $\alpha$, it is necessary to determine the off-resonant point $P$. This point $P=1$ is an invariant quantity of the resonator model. It is independent of the resonance frequency, the quality factor and the impedance mismatch. Its position in the complex plane $P'$ is only altered by the prefactors $a$ and $\alpha$ of the environment. Thus, knowing its position determines these two remaining quantities. In order to find that point, one needs to know the resonance frequency. This is done fitting a circle to the data and translating its center to the origin, see Fig.~\ref{fig:normalization}(b). Then, a phase vs. frequency fit is performed shown in Fig.~\ref{fig:normalization}(c), which yields the resonance frequency, the total quality factor $Q_l$ and the offset phase $\theta_0$ (see Ref.~[\onlinecite{GaoPhD2008}])
\begin{equation}
\theta(f)=\theta_0+2 \arctan\left(2 Qr \left[1-\frac{f}{f_r}\right] \right)\,.
\label{eq:phasevsfreq}
\end{equation}
From $\theta_0$ and the center position of the circle, the complex position of the transformed off-resonant point $P'$ is found via geometric relations
\begin{equation}
\tilde{P} = x_c+r_0 \cos\left(\beta \right) + i \left[y_c+r_0\sin\left( \beta \right)\right]\,.
\label{eq:offrespoint}
\end{equation}
Here, $ \beta = \left(\theta_0+\pi\right)\, \text{mod}\, \pi$ denotes the angular position of the off-resonant point $P'$ of the translated circle, which is rotated by $\pi$ from the resonance point, $\beta \in (-\pi,\pi)$. The absolute value of $P'$ gives the amplitude scaling factor $a$ and $\arg (\tilde{P} )$ yields the phase offset $\alpha$. Figure \ref{fig:normalization}(d) displays the final transformation of the circle to its canonical position.

\section{Implementation}
In a real experiment, one would use the described algorithm as follows. First, a high resolution scan of the resonator $S_{21}(f)$ is taken. This is usually done at high power and at low IF bandwidth such that the signal-to-noise ratio (SNR) is larger than 100. One should take a sufficient number of points evenly spaced around the resonance and distributed over a frequency span of approximately four 3~dB bandwidths. Here, the 3~dB bandwidth is defined by the loaded quality factor and the resonance frequency $\Delta f_\text{3dB}=f_r/Q_l$. The automatic determination of the prefactors $a$, $\alpha$ and $\tau$ is very accurate in this case. For the subsequent measurements, all measurement data is normalized by these prefactors before performing the actual circle fit method. This normalization rotates the resonance circle in the canonical position, see Fig.~\ref{fig:normalization}(a) to (d), such that from now on only 4 fitting parameters remain to be determined. This allows for maintaining a good accuracy of the fit even at low SNRs, which reduces the measurement time. Furthermore, as shown below, much less points are needed in the subsequent fits, which speeds up the measurement further.\\
The calibrated data are first fitted to a circle of the $\text{Im}\{S_{21}\}$ vs. $\text{Re}\{S_{21}\}$ data determining the center $(x_c,y_c)$ and radius $r_0$. The impedance mismatch is directly determined by $\phi_0=-\arcsin(y_c/r_0)$. Then, the circle is translated to the origin and a phase vs. frequency fit is performed, see Eq.~(\ref{eq:phasevsfreq}). Here, the initial value for $\theta_0=\phi_0+\pi\, $mod$\, \pi$ is calculated. This gives $Q_l$ and $f_r$. Since the circle is already in its canonical form, we can calculate \cite{Khalil_12} the complex coupling quality factor $Q_c=Q_l/(2r_0 \exp(-i\phi_0))$. The internal quality factor is given by equation $Q_i^{-1} = Q_l^{-1} - \text{Re}\left\{ Q_c^{-1} \right\}$. The errors of the $\lambda$ individual fitting parameters $\xi_i$ are conveniently calculated by first numerically determining the Jacobian matrix $J=-\partial \chi_i/ \partial \xi_j$ with $\chi_i = |y_i - S_{21}(f_i,\xi)|$. Then, the covariance matrix is given by $\sigma^2=\chi^2/(N-\lambda)(J^TJ)^{-1}$ and the errors are the square roots of its diagonal elements. An implementation of the algorithm is available on our website\cite{sourcecode}.

\section{Test of the algorithm}
In this section, we investigate the fit's behavior in the presence of different noise levels. For this purpose, we generate artificial $S_{21}$ data and add a controlled amount of noise. This is done in such a way that the radius $r_0$ of each point on the resonance circle is altered by a Gaussian distributed random number. The distribution has its mean value at $r_0$ and a width of $\sigma=r_0/\text{SNR}$, where SNR denotes the desired signal to noise ratio of the generated data.
The signal to noise ratio is defined as follows \cite{Petersan_JAP98}
\begin{equation}
\text{SNR} = \frac{r_0}{\sigma_r} = \frac{r_0}{\sqrt{\frac{1}{N-1}\sum_{i=1}^N{\left(r_i-r_0\right)}}}\,.
\label{eq:SNR}
\end{equation}
Here, $r_0$ denotes the radius of the circle, N is the number of data points and $r_i=\sqrt{\left(x_i-x_c\right)^2+\left(y_i-y_c\right)^2}$ with $\left(x_c,y_c\right)$ denoting the center of the circle. Only in the case of artificially generated data, the SNR can be computed exactly. For real data, the center and radius have to be fitted first and are subject to errors as well. However, in most cases, the obtained values should be accurate enough to determine the SNR.\\
\begin{figure}[t]
\includegraphics[width=\columnwidth]{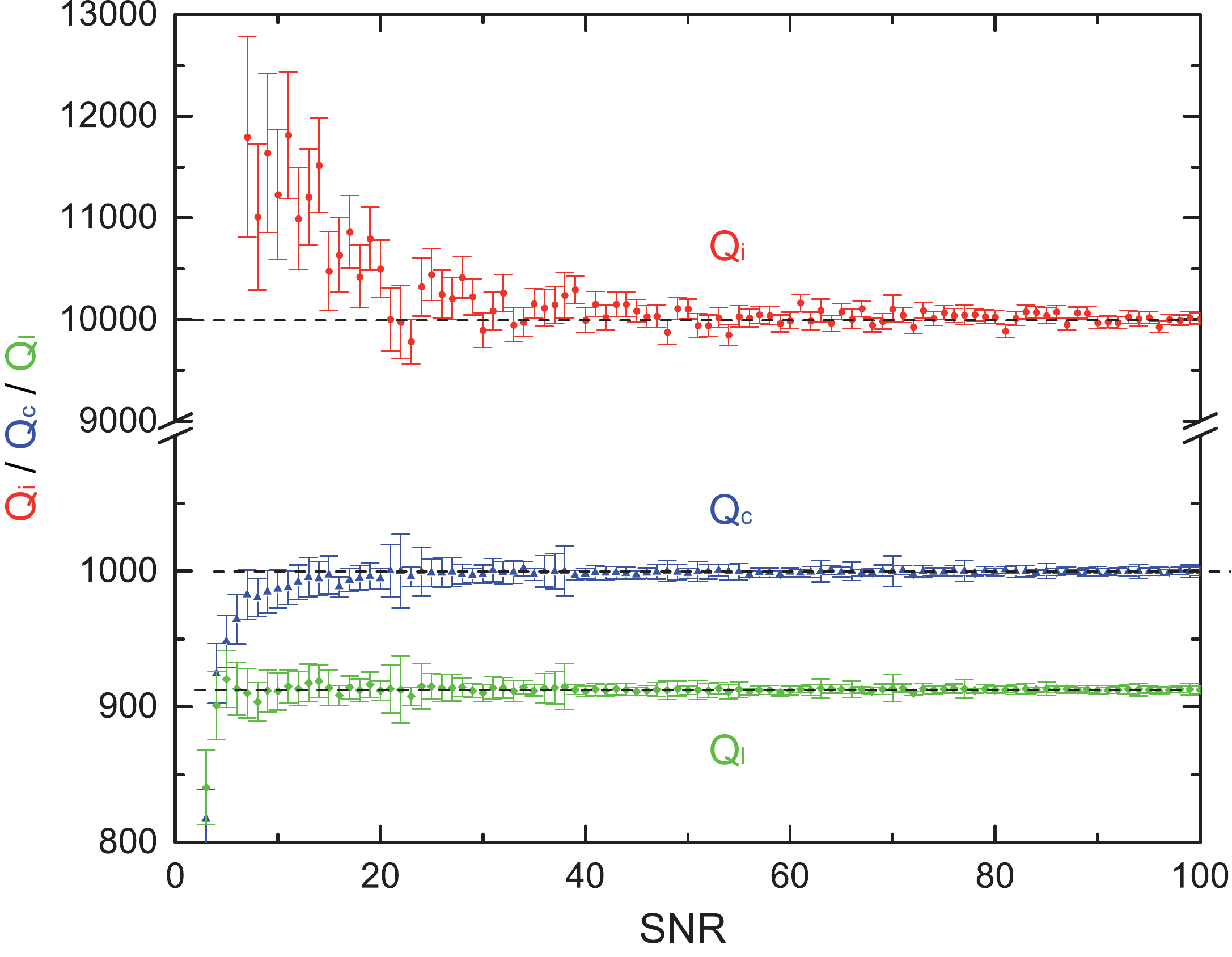}%
\caption{(color online) Internal $Q_i$, coupling $Q_c$ and total quality factor $Q_l$ as a function of the signal to noise ratio. The internal quality factor is already diameter corrected. The error bars denote the statistical error, which is estimated by the algorithm, and the dashed lines indicated the respective true values. Above SNR=40, the fitted values are very stable, and down to SNR=20, the fit of $Q_i$ remains robust.}%
\label{fig:resFig1}%
\end{figure}
The chosen parameters of the generated data are typical for superconducting resonators: $Q_c=10^3$, $Q_i=10^4$, $f_r=5~$GHz, $\phi=0.03\,\pi$. For the test of the calibration algorithm, we set the prefactors as follows: $\tau=50\,$ns, $a=0.1$, $\alpha=0.4\,\pi$. Figure~\ref{fig:resFig1} shows the fit results 
as a function of the signal to noise ratio from 1 to 100, assuming a properly calibrated 
measurement. Interestingly, even in the low SNR regime, the algorithm performs well.
\begin{figure}[!ht]
\includegraphics[width=\columnwidth]{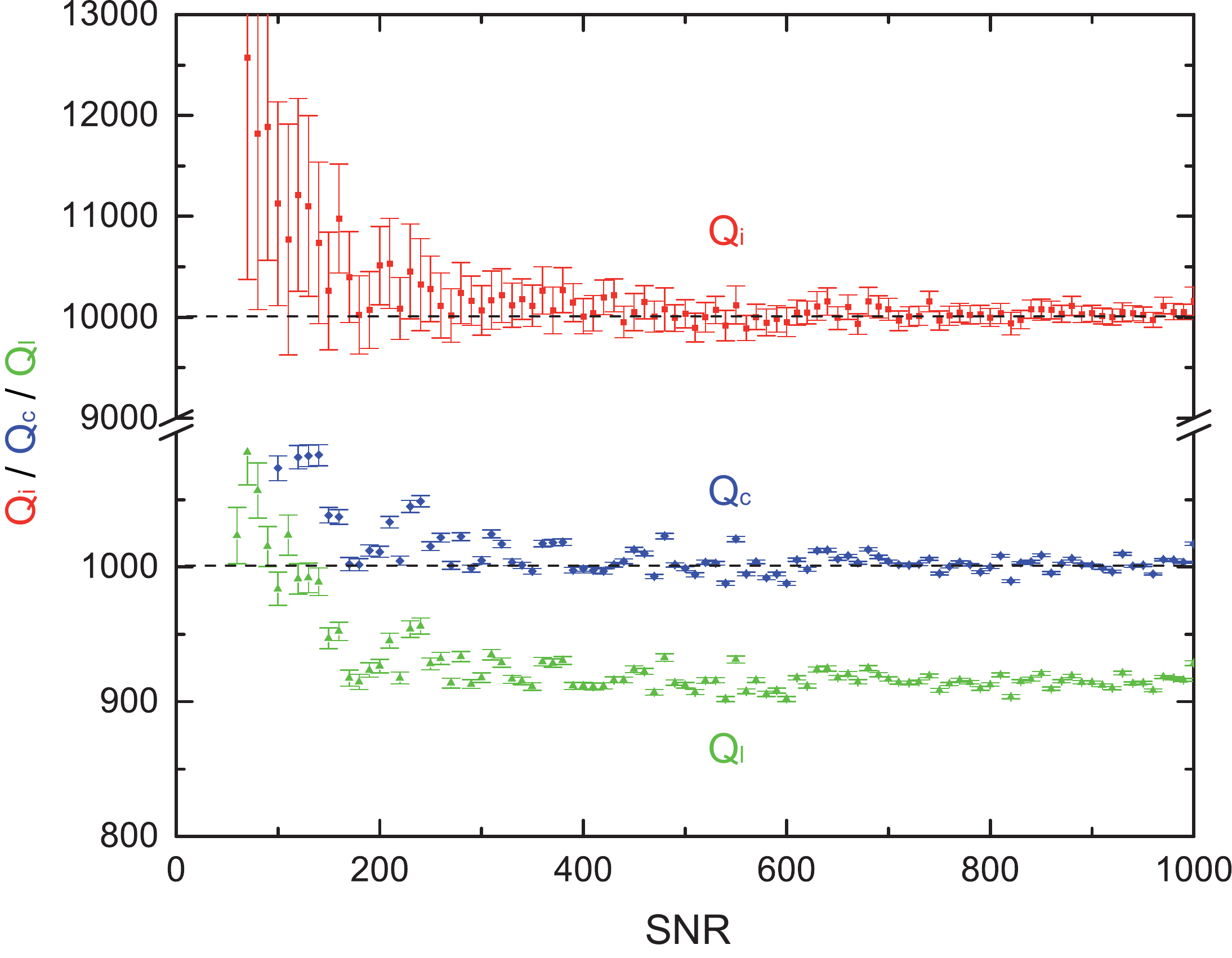}%
\caption{(color online) Internal, coupling and total quality factors as a function of the signal to noise ratio. The dashed lines indicate the true $Q_i$ and $Q_c$ values. Here, the raw data contain additionally an arbitrary cable delay, start phase and amplitude, which needs to be automatically corrected for in advance. Thus, this complex procedure works only for high quality data with SNR larger than 300, for the given parameters.}%
\label{fig:resFig2a}%
\end{figure}

\begin{figure}[!ht]
\includegraphics[width=\columnwidth]{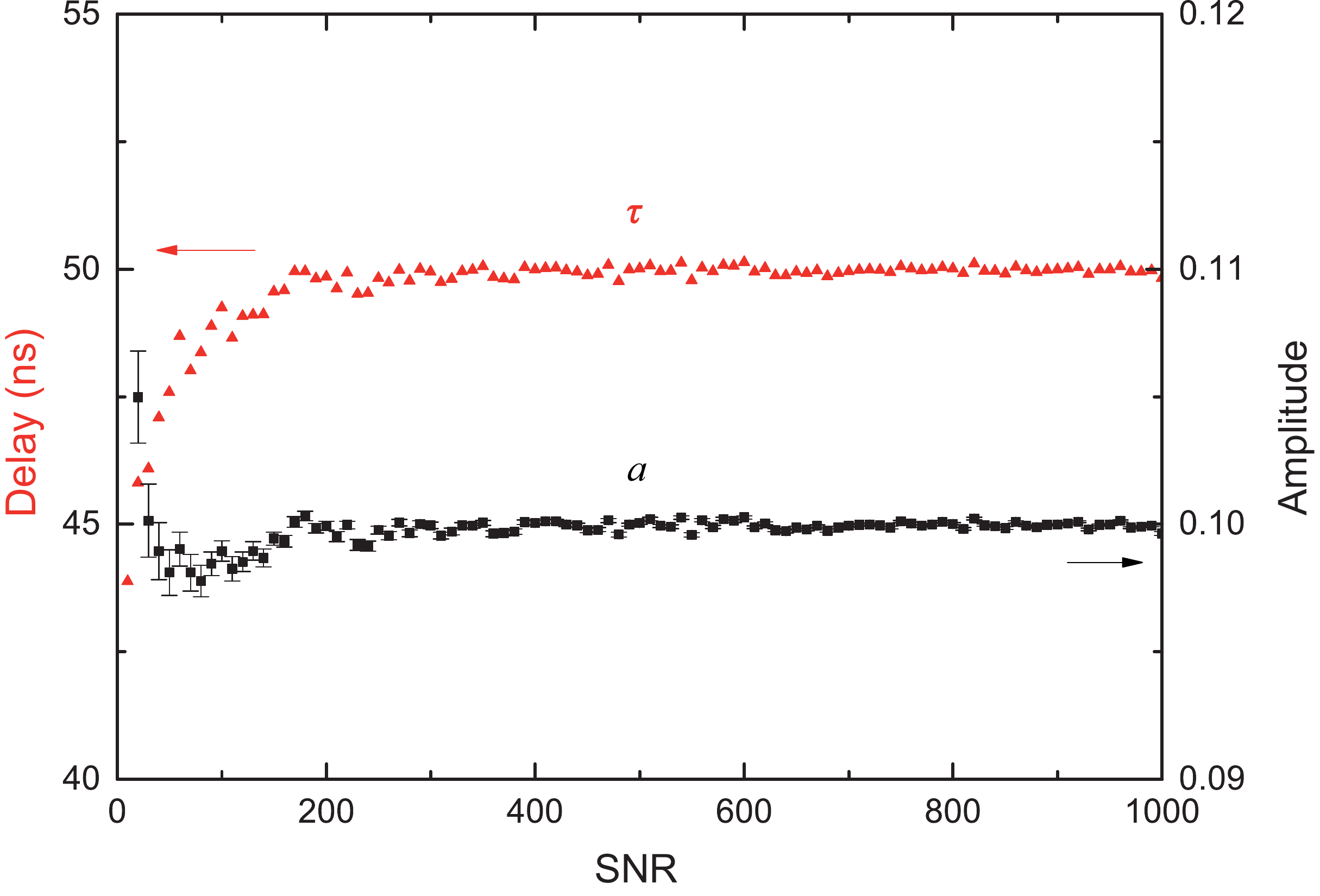}%
\caption{(color online) Fitted cable delay and amplitude as a function the SNR. }%
\label{fig:resFig2b}%
\end{figure}

Next, Figs.~\ref{fig:resFig2a} and \ref{fig:resFig2b} show the performance of the calibration algorithm as a function of the SNR from 10 to 1000. Figure \ref{fig:resFig2b} shows the results of the automatic cable delay and amplitude calibration procedure. As expected, the calibration works only for large SNR values. Below a SNR of 200, the calibration becomes inaccurate and, therefore, the determination of the resonator's parameters is not reliable anymore. But this is not an obstacle because a calibration is typically run only once at large SNR for a given experiment.

\begin{figure}[!ht]
\includegraphics[width=\columnwidth]{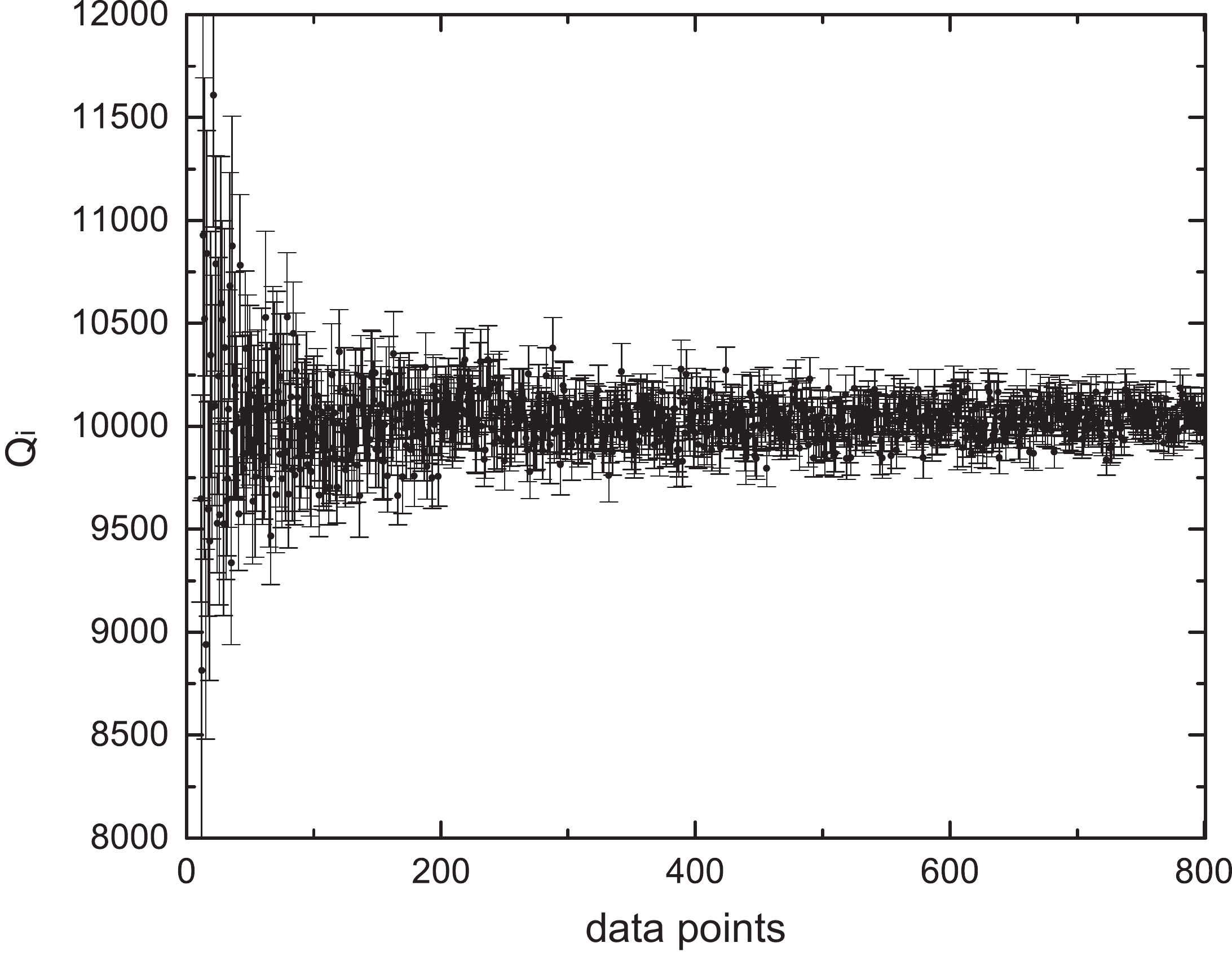}%
\caption{(color online) Internal quality factor as a function of the number of recorded points. Above 200 points, this fit appears to be only limited by the SNR, which is set to 65 in this case.}%
\label{fig:resFig3}%
\end{figure}

Finally, the performance in the limit of very few points is tested. For many experimental settings it may be convenient to take only few data points but with a larger SNR. Figure \ref{fig:resFig3} shows the performance of the calibrated algorithm as a function of the number of points from 11 to 801 for a fixed SNR of 65.
One can see that at least 200 or more points are necessary for a reliable determination of the internal quality factor at the given SNR level.

\section{Conclusion}
The presented method offers a fast and accurate analysis of noisy complex scattering data of microwave resonators. At sufficient signal-to-noise ratio, the method relies on complex VNA data without any need of calibration. As a result, all six parameters of the generalized resonator problem are determined automatically. These are internal and loaded quality factors, the resonance frequency, and an impedance mismatch parameter. In addition, the contributions of the measurement circuit as well as the cable delay are accounted for; given by an arbitrary amplitude and phase factors. In the presence of strong noise, the calibrated resonator fit algorithm proves to be very robust even at $\text{SNR}<20$. Finally, the presented fit algorithm outperforms existing methods in terms of robustness and speed because it uses a non-iterative fit for the determination of the circle parameters, i.e. the circle fit converges immediately without the need of providing start values. The method allows to analyze calibrated resonator data in real time, which is required for a large range of applications of superconducting microwave resonators.

\section{Acknowledgments}
We thank J. Braum\"uller for the critical reading of the manuscript, and we gratefully acknowledge valuable discussions with D.~Pappas, M.~Vissers and J.~Gao. S.~P.~acknowledges financial support by the LGF of Baden-W\"{u}rttemberg. This work was supported in part by the DFG, the BMBF program "Quantum communications" through the project QUIMP and the Ministry of Education and Science of the Russian Federation under contract no.~11.G34.31.0062.

\bibliographystyle{apsrev4-1}
\bibliography{circlebib}

\end{document}